# Magneto-optical imaging of magnetic flux patterns in superconducting films with antidots

V.V. Yurchenko[a,b*], R. Wördenweber[c], Yu. M. Galperin[a], D. V. Shantsev[a],

J.I. Vestgården[a] and T.H. Johansen[a]

[a] *Department of Physics, University of Oslo, P.O.Box 1048, Blindern, 0316 Oslo, Norway*

[b] *Institute of Physics ASCR, Na Slovance 2, 18221, Prague, Czech Republic*

[c] *Institut für Schichten und Grenzflächen (ISG-2) Forschungszentrum Jülich GmbH, D-52425, Jülich*

**Abstract**

We present the results of experiments on visualization of magnetic flux distribution and its dynamics in high-temperature superconductors with artificial defects. High-$T_c$ superconductor thin films were equipped with a special arrangement of antidots in order to separate the streams of magnetic flux moving in (or out of) the sample. A possibility to alter the direction and depth of flux penetration is clearly demonstrated by means of magneto-optical imaging. The resolution was sufficient for observation of flux in particular antidots, which allows more detailed dynamic analysis of such systems.

## 1. Introduction

Lithographically produced holes in superconducting thin films – antidots – have been a subject of numerous experimental and theoretical investigations for several decades [1-11]. Discovery of new classes of superconducting materials and advances in lithographic techniques revitalized the topic and stimulated a new wave of design and investigations of artificial defects. This brought many ideas about utilization of antidots in superconducting devices [5-8]. Those include vortex ratchets, which employ the capability of antidot arrays of guiding magnetic flux in selected directions.

Earlier transport and magnetization studies of conventional and high-$T_c$ superconductor thin films with regular arrays of antidots prove that under some conditions antidots may act as sites of strong pinning and lead to matching effects [10-11]. There is also evidence that under other conditions arrays of antidots reduce the barrier for flux propagation [5]. In spite of many experimental indications and theoretical predictions, which often produce ambiguous or even controversial interpretations, a general picture of field distributions and mechanisms of flux motion in superconductors with antidots still remain poorly understood. *Direct* visualization of the magnetic flux and its dynamics would help to shed light on these issues and eventually improve design of fluxonic devices.

Magneto-optical (MO) imaging offers a good combination of high magnetic and spatial resolution at relatively high acquisition rates. Another advantage of this method is that simply by changing the magnification of objective lenses (see experimental details below) one can vary the area of interest, zooming from the overall view of the sample into the details of field distribution inside the micrometer holes. Recently, simple 2D periodic arrays of antidots were investigated by means of MO imaging, and several papers were published [12-14].

In this article, we present MO images of a high-$T_c$ superconducting film equipped with line segments of antidots of different alignment with respect to the film edge. In particular, we investigate a ratchet configuration suggested in [5], where the idea is to let the Lorentz force guide part of the flux into dead ends. We analyze the observed field distributions and supplement the analysis with results of computer simulations.

[*] Corresponding author. Tel.: +47 228 54125; fax: +47 228 56422; e-mail: vitaliy.yurchenko@fys.uio.no.

## 2. Experimental details.

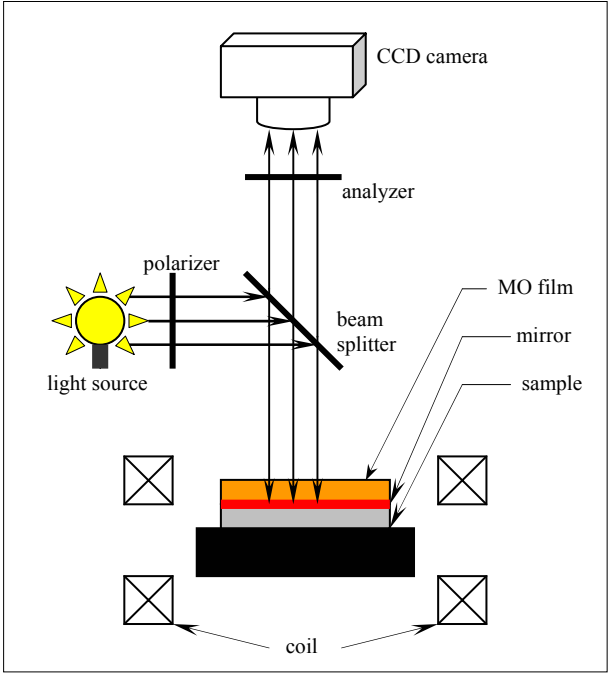

Fig. 1.Schematic drawing of the MO imaging setup.

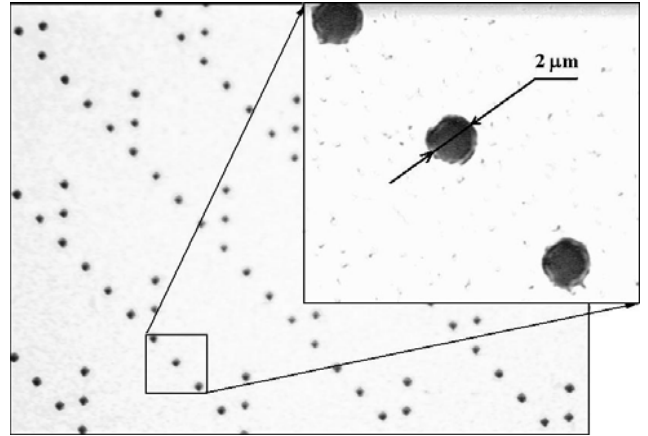

Fig. 2. Optical image of the ratchet sample. The inset shows an enlarged image. The distance between antidots is 8 μm.

*2.1. Experimental technique*

MO imaging is based on the Faraday effect, i.e. rotation of the polarization plane of incident linearly polarized light in materials with longitudinal optical birefringence in magnetic field. As Faraday active sensor we use a bismuth-doped ferrite garnet film with in-plane magnetization. The garnet film was grown by liquid phase epitaxy on gadolinium gallium garnet substrate. The Faraday rotation depends on the film thickness $l$, the magnetic field component $H$ parallel to the incident light and a material parameter $V$, the so-called Verdet constant. At small field values the Faraday rotation angle is given by:

$$\theta = VlH \qquad (1)$$

Our MO imaging setup (see Fig.1) consists of a mercury lamp light source, a Leica polarizing microscope, an Oxford optical cryostat, a pair of coils and a 12-bit CCD (charge-coupled device) camera.

For mapping the field over the surface of a superconductor the MO layer is brought into close contact with the sample. To double the Faraday effect and achieve better reflectivity a thin aluminum mirror layer is sputtered on the MO film. With our setup the spatial resolution goes down to approximately 0,8 μm, which makes it possible to visualize a few flux quanta localized at an antidot. Due to a high dynamic response of the MO film (less than $10^{-9}$ seconds) this method can be used for real time magnetic visualization.

*2.2. Samples*

One of the key properties of superconducting thin film devices like ratchets, is their capability of guiding magnetic flux in selected directions. Changing the direction of vortex motion can be realized by creating spatially inhomo-geneous pinning potential or by introducing extended defects, such as slits or large holes, which would locally modify the current flow and, consequently, the direction of the Lorentz force. For the present work a 150 nm thick $YBa_2Cu_3O_{7-\delta}$ thin film sample was produced by magnetron sputtering. The sample has the shape of a long strip with a width of 500 μm. Holes with radii of 1 μm were produced by optical lithography and ion beam etching. The antidots, separated by 8 μm, were arranged in the ratchet configuration suggested by Wördenweber *et al.*[5] This consists of lines of equidistant antidots forming a 45 degree angle with the edge of the superconducting strip, see Fig.2. In addition, short perpendicular segments of antidots were introduced in order to separate the flux flow and guide some flux into the dead-ends.

## 3. Results and discussion.

Shown in Fig. 3 are MO images of the magnetic flux distribution in the sample at three consecutive values of external magnetic field. The field was slowly ramped up after cooling the sample to 4 K in zero field. In our experiments we used crossed polarizers, hence, the image brightness represents the magnitude of the local magnetic induction.

It is clearly seen in Fig. 3 that the flux density in the antidots is significantly higher than in the surrounding superconductor. The effective depth of flux penetration along the lines of holes is also much greater than that between them (or in intact films). Hence, we conclude that the lines of antidots of radius $R \approx 1$ μm *facilitate* propagation of magnetic flux rather than acting as additional pinning sites.

The resolution achieved in our experiments allows a detailed analysis of the flux distribution and dynamics in and around the antidots. At low applied fields, see Fig. 3(a), only the holes next to the edge are populated with flux. At elevated fields, (b) and (c), the flux is transferred also to more distant holes while the occupation number of the already populated holes increases. It seems that every populated hole becomes a link in the chain transferring the flux: it receives flux from its neighbour and passes it on to the next hole deeper into the sample.

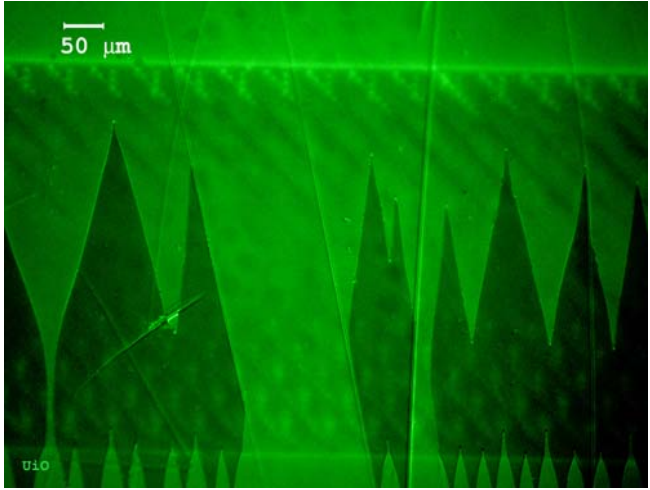

a)

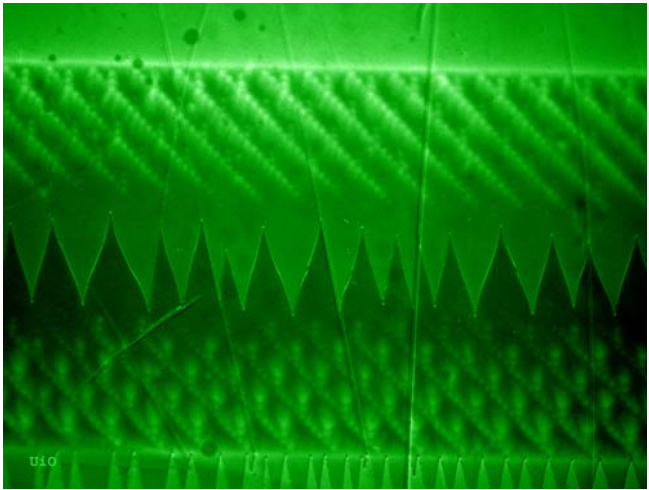

b)

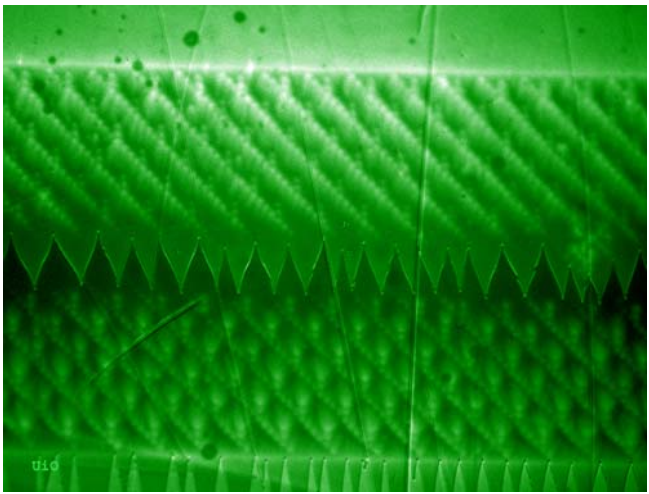

c)

Fig. 3. MO images of flux distributions in the superconducting strip in an increasing magnetic field after zero-field-cooling to $T = 4$ K. The images (a)-(c) were taken at applied fields of $B_a = 1$ mT, 4 mT and 6 mT, respectively. The zigzag lines are domain walls in MO garnet film and should be disregarded

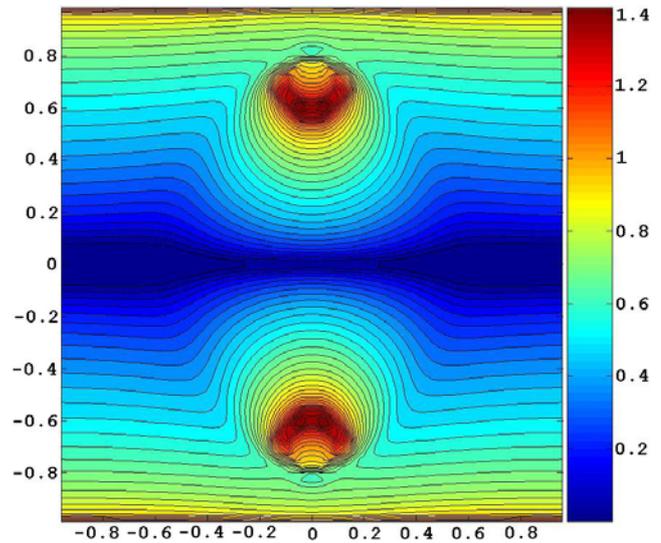

Fig. 4. Distribution of magnetic field in a strip with two holes obtained from computer simulations (arbitrary units). Bends in current flow profile cause parabolic distribution of magnetic field.

Meanwhile, some flux leaving antidots penetrates into the superconducting area ahead of it forming parabolic features seen in Fig. 3, and more clearly in Fig. 5. These parabolas are especially pronounced at the ends of vertical lines – the "dead-ends" (Fig. 5, right). They originate from the last antidot in a row which does not have a possibility to pass the flux on to the next hole. Hence all the flux guided to the dead-end has to accumulate there or enters the superconductor.

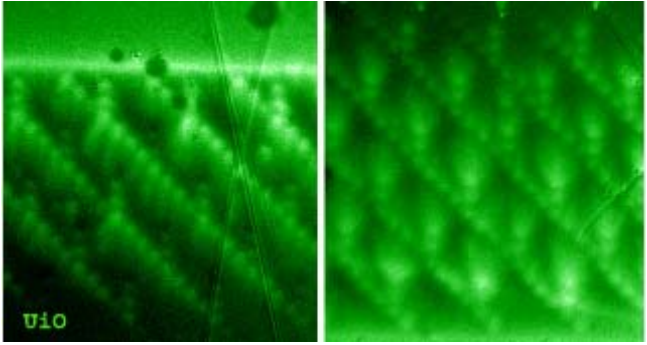

Fig. 5. Enlarged MO image of superconductor strip with antidot ratchet. *p= 8µm, T=4K, B=4mT*. Top (left) and bottom (right) edges of the strip.

Fig. 5 (left) shows another specific type of antidots where the two lines – diagonal and vertical – merge together. Hence, the flux is guided to them via two channels, but can move on only via one diagonal channel. As a result, these holes become overpopulated and some extra flux has to penetrate the superconductor. Indeed, we can see that these antidotes serve as the nucleation sites for relatively large parabolic features on the images. At high fields a whole line of antidots represents a cascade of parabolas with the centres in the holes.

The origin of parabolic flux distribution is the meandering of the current around the holes. The parabolas seen in the MO images separate domains of different current orientations (they are often called d-lines) [15,16]. This is illustrated by Fig. 4, which shows the result of computer simulations of flux penetration in a thin superconducting strip with current-voltage curve $E \sim j^n$, n=15 based on the formalism developed by Brandt [17, 18]. The strip contains two holes which give rise to parabolic features in the flux pattern (the flux density is constant along the lines).

Finally, it is known that an antidote cannot carry more flux quanta than a certain saturation number [1]. In the present experiment we never reach this number since the flux density in the holes keep increasing while we ramp the applied field up to the maximal value of 10 mT.

## 4. Conclusions

We demonstrated by means of magneto-optics that in YBCO thin films with 1D arrays of artificially patterned holes the propagation of magnetic flux is significantly facilitated along the lines of artificial defects.

Every hole transfers some flux to the next hole and some - to the superconductor ahead of it. The flux in the superconductor acquires a parabolic pattern with centers of parabolas sitting in the antidots. An additional evidence of flux guidance by the hole arrays comes from the observation of most pronounced parabolas at the antidots fed by two flux channels or situated at the dead-ends, where the further flux transfer is impossible.

**Acknowledgements**

The work was supported by the Research Council of Norway, Grant. No. 158518/431 (NANOMAT). Discussions with Dr. E.K. Hollmann, Prof. A.V. Bobyl, B. Rosewig and Å.A.F. Olsen are acknowledged.